\documentstyle[aps,prc,preprint]{revtex}
\draft
\title{Energy dependence of the isomeric cross section ratio in the
	$^{58}$Ni$(n,p)^{58}$Co$^{m,g}$ reactions}
\author{V. Avrigeanu$^{1,2}$, S. Sud\' ar$^2$, Cs. M. Buczk\' o$^2$, 
J. Csikai$^2$, A.A. Filatenkov$^3$, S.V. Chuvaev$^{2,3}$, 
R. D\' oczi$^2$, V. Semkova$^{2,4}$, and V.A. Zelenetsky$^5$}
\address{$^1$Institute for Physics and Nuclear Engineering, 
		P.O. Box MG-6, 76900 Bucharest, Romania\\
$^2$Institut of Experimental Physics, Kossuth University, 
		H-4001 Debrecen, Postfach 105, Hungary\\
$^3$V.G. Khlopin Radium Institute, 2nd Murinski Ave. 28, 
		St. Petersburg 194021, Russia\\
$^4$Institute of Nuclear Research and Nuclear Energy, 
		1784 Sofia, Bulgaria\\
$^5$Institute of Atomic Energetics, Studgorodok 1, 249020 Obninsk, 
		Russia}

\begin{document}
\maketitle

\begin{abstract}
Excitation functions of the $^{58}$Ni$(n,p)^{58}$Co$^{m,g}$ reactions 
were measured in the energy range from 2 to 15 MeV. The energy 
dependence of the isomeric cross-section ratio 
$R=\sigma_m/(\sigma_m+\sigma_g)$ is deduced from the measured data. 
The shape and magnitude of the $R(E_n)$ function are described by 
model calculations using a consistent parameter set. Questions of the 
input level scheme were solved based on the accurate isomeric ratio 
measured at low energy region.\\
\end{abstract}

%\section{INTRODUCTION}

A previous survey of isomeric cross sections pointed out the practical 
significance of these data in fusion reactor technology as well as in 
the production of medically important radioisotopes \cite{qaim94}. 
Furthermore, isomeric cross-section ratios are of basic interest for 
the analysis of nuclear-spin effect within the formation of isomeric 
states in nucleon \cite{qaim94} and heavy-ion \cite{dig90} induced 
reactions. An illustrative case for the experimental difficulties and 
the contradictory calculated results is the $R(E_n)$ function in the 
case of $^{58}$Ni$(n,p)^{58}$Co$^{m,g}$ reactions. Recent measurements 
and model-calculations \cite{buczko95,sudar96} on the $R(E_n)$ 
function have required further investigations to solve the discrepancy 
in the data \cite{avr97}.

%\section{EXPERIMENTAL PROCEDURE}

Metallic foils made of enriched $^{58}$Ni and natural Ni were 
irradiated with neutrons produced via the $^2$H$(d,n)^3$He and 
$^3$H$(d,n)^4$He reactions around 2.5 MeV and 14 MeV, respectively. 
Irradiations were carried out at KRI (St. Petersburg) and IEP 
(Debrecen) in  scattering free arrangements. Neutron flux variation in 
time and at the sample position was measured in KRI by two independent 
scintillation detectors. The neutron energy distribution inside 
samples was calculated by a code taking into account the real 
experimental conditions such as the  size of the beam, the solid angle 
for the sample, the slowing down and scattering of D$^+$ beam in the 
target. 

Usually, the foils were irradiated for 5 h with $D-D$ and 1 h with 
$D-T$ neutrons, respectively. Detection of the 810.8 keV gamma-line 
was started immediately after irradiation using several detectors 
simultaneously. Each sample was measured continuously for 2--3 days 
and the measured spectra were saved in every one or two hours. The 
intensity of the 810.8 keV gamma-line from the decay of the $^{58g}$Co 
ground state (T$_{1/2}$ = 70.92 d) populated  directly and also by the 
decay of the $^{58m}$Co isomeric state (T$_{1/2}$ = 9.15 h) was 
determined as a function of time. These data were compared with the 
calculated decay curve containing both the $\sigma_m$ and $\sigma_g$ 
parameters. The best values of isomeric ratios were obtained using the 
weighted least-squares method for the adjustment of the calculated 
curves to the experimental points. 

The uncertainty in the isomeric ratio indicated in Table I was deduced 
from repeated measurements.

The quasi-monoenergetic neutrons in the 5.38 -- 12.38 MeV range were
produced by the MGC-20 cyclotron of ATOMKI (Debrecen) using D$_2$ gas 
target. The activities of samples were determined by HPGe, NaI and 
Ge(Li) detectors.  Details of the experimental procedures have been 
published elsewhere \cite{buczko95,filat97,filat971}. The measured 
data are given in Table I. The data points of the $R(E_n)$ function
between 5 and 10 MeV have been deduced from the ($\sigma_m+\sigma_g$) 
and $\sigma_m$ values measured in KFA (J\" ulich) by the Co X-rays
\cite{sudar91} emitted in the decay of $^{58}$Co$^m$.

%\section{NUCLEAR MODEL CALCULATIONS}

Preequilibrium-emission (PE) and statistical model calculations were 
carried out by using the computer code STAPRE-H95 \cite{sh95}. The PE 
processes have been described by means of the Geometry-Dependent 
Hybrid (GDH) model including the angular momentum and parity 
conservation \cite{avr88,avr90} which leads to an enhanced PE from 
higher spin composite-system states and higher orbital momenta in the 
emergent channels. These aspects are particularly important to our
better understanding of the isomeric cross sections \cite{bissem80}). 
A consistent parameter set, established or validated by means of 
different types of independent experimental data \cite{avr90,avr94}
was involved in the GDH calculations. The corresponding discrete level 
data and level density parameters of the back--shifted Fermi gas 
(BSFG) model are given in Table II. Particular optical model potential 
(OMP) parameter sets have been used for neutrons on $^{58}$Ni 
\cite{chiba90} and $^{59}$Co \cite{smith88}.

The calculated $R(E_n)$ excitation function obtained by using the 
evaluated value \cite{peker90} for the branching ratio of the 52.8 
keV$\rightarrow$24.9 keV transition is shown in Fig. 1(a). It was 
found that the calculated cross sections for the 
$^{58}$Ni$(n,p)^{58}$Co$^m$ reaction is lower with $\sim$ 35\% than 
the experimental data. As shown in Fig. 1(b) similar behavior can be 
observed for the $^{59}$Co$(n,2n)^{58}$Co$^m$ reaction, in which the 
same $^{58}$Co residual nucleus is produced.

%\section{RESULTS AND DISCUSSION}

In order to study the accuracy of the statistical-model calculation,
the advantage of the accurate $R$-values measured at low energy region
was considered \cite{filat97}. In this case only the statistical 
population of the lowest few discrete levels and the corresponding 
decay scheme are important. As shown in Fig. 1(a) by using only the 
ground and isomeric states the lack of agreement between the 
experiment and model prediction is gradually corrected for increasing 
incident energies. This has restricted the possible sources of the 
underpredicted values to the decay scheme of the very low-lying 
levels. It was found that a branching ratio of (75$\pm$5)\% for the 
52.8 keV$\rightarrow$24.9 keV transition brings into agreement the 
calculated and these particular experimental data. At the same time, 
the energy dependence of the isomeric cross-section ratio for the 
$^{59}$Co$(n,2n)^{58}$Co$^{m,g}$ reaction as well as the excitation 
functions of both the $^{58}$Ni$(n,p)^{58}$Co$^{m,g}$ and
$^{59}$Co$(n,2n)^{58}$Co$^{m,g}$ reactions are also well reproduced,
as shown in Figs. 1(c) and 1(d).

The calculated $R(E_n)$ excitation function based on the 
above-mentioned assumption has been tested by an additional analysis 
of the $^{58}$Co level scheme effect. It should be noted that the most
recent evaluation of the level schemes for nuclei with atomic mass 
$A$=58 \cite{bhat97} became available when this work was mainly 
carried out \cite{avr97}. However, the number of adopted levels of 
$^{58}$Co corresponding to the excitation energy considered in this
work (Table II) was decreased by only one, while the re-evaluated 
branching ratios using the same experimental data base should be
changed also for a single level. In order to check our previous 
considerations \cite{avr97} we have used the new evaluation.

The shape of the compound nucleus angular momentum ($J$) distribution 
given within the statistical model by the neutron OMP \cite{chiba90} 
is shown in Fig. 2(a). At the lowest incident energies it is bell-like 
and nearly symmetric at around the average $\overline{J}$ value (which 
is, e.g., $\sim$2 $\hbar$ around $E_n$=4 MeV) and becomes nearly 
triangular above $\sim$10 MeV. This distribution may explain the role 
of various assumptions involved in the case of the adopted levels 
\cite{peker90} which have no spin assignment. The question is less 
severe for such levels which have at least a known branching ratio. 
The spin values considered for them in the present calculations, 
marked additionally in Fig. 2(b), are confirmed for two levels by the 
superseding evaluation \cite{bhat97} and differ by one unit for a 
third level. The other case happens for three levels, the guidance by 
the level scheme for $^{56}$Co (with a shell-model configuration 
having not three but one neutron in the $p_{3/2}$ shell) being useful 
only for one of them considered as a 0$^+$ level. The yrast plot has 
suggested the assignment 4$^+$ for the other two levels of this kind, 
and we have assumed for them an uniform decay to the ground and 
isomeric states. This main choice leads to the solid curve in Fig. 
1(a), while we have alternatively considered also both these 4$^+$ 
levels populating either the ground state (lower dashed curve) or the 
isomeric state (higher dashed curve). Any other option, e.g. both 
levels being 0$^+$ or 6$^+$ and decaying only to the g.s. and, 
respectively, isomeric state, provides $R(E_n)$ excitation functions 
between the above-mentioned limits. Therefore, it was found 
\cite{avr97} that an uncertainty of about 10\% in the isomeric 
cross-section ratio comes from various spin assignments for only two 
of the 29 discrete levels of the product nucleus.

The final use of the 28 adopted levels up to 1.555 MeV excitation 
energy \cite{bhat97} provides an effective check of the above 
comments. The adoption of the re-evaluated level and decay schemes
\cite{bhat97} leads to changes of the calculated $R(E_n)$ values from 
$\sim$3\%, around the incident energy of 2 MeV, to $\sim$0.3\% around 
14 MeV. On the other hand, the only change of the branching ratios for 
the third excited level \cite{bhat97} has an enhanced effect on the 
branching-ratio value of (85$\pm$5)\% for the 
52.8 keV$\rightarrow$24.9 keV transition which makes possible the 
agreement between the experimental $R(E_n)$ data at 2--3 MeV and the 
calculated results [the solid curve in Fig. 3(a)]. The greatest 
difference with respect to the calculated $R(E_n)$ excitation function 
by using the previous level scheme \cite{peker90}, shown by the solid 
curve in Fig. 1(a) and dashed curve in Fig. 3(a), is just within the 
limit of 10\% discussed previously.

The meaning of the level schemes of the both residual and competitive 
reaction channels, for (i) the slope of the calculated $R(E_n)$ 
excitation function, and (ii) some "structure" present at the lowest 
energies (Fig. 3) was analysed. Actually, this (n,p) reaction on an 
even-even neutron-deficient target nucleus, with a small but positive 
$Q$-value, is a rather particular case. The competition between the 
even-even $^{58}$Ni and the doubly odd $^{58}$Co residual nuclei 
should be also carefully considered, especially at lower energies. The 
analysis illustrated in Fig. 3 makes possible to identify the effect
of each of the target-nucleus lowest discrete levels as well as, 
beyond this behavior, a trend similar to the heavy-ion induced 
reactions at sub-barrier energies \cite{dig90} i.e. rather constant 
$R$ values just above the reaction threshold. It may also be supported 
the conclusion \cite{dig90} that deficient discrete-level schemes used
in earlier analyses require a nuclear moment of inertia lower -- 
typically one-half -- as the rigid-body value $I_r$ (with reduced 
radius $r_0$=1.25 fm) in order to reproduce the measured $R$ data. 
Actually, it was shown that there is no reduction of the effective 
moment of inertia below the rigid body value, e.g., by Fisher {\it et 
al.} \cite{fis84} through study of the spin cut--off parameters for 
$^{53}$Cr and $^{57}$Fe derived from analysis of neutron--induced 
reactions at 14.1 MeV; this result was next successfully used by the 
same group of IRK--Vienna for description of the all neutron reactions 
on $^{58}$Ni up to 20 MeV \cite{pav85}, and in subsequent calculations 
in the range $A$=46-64 \cite{avr90,avr94}.

Calculation of the $R(E_n)$ excitation function has been carried out 
by using the assumption of the one--half rigid body value for the 
nuclear moment of inertia. The corresponding other two BSFG parameters 
have been obtained by the fit of the same discrete level data (Table 
II). However, the calculated isomeric cross--section ratio does not 
depend on the nuclear level density for incident energies lower than 
$\sim$3.5 MeV. Above this energy up to 15 MeV the calculated values by
using 0.5$I_r$ are lower, and can achieve about 12\% at the highest 
energy, with respect to the solid curve in Fig. 1(a). Except the 
better agreement with the three experimental data around 12 MeV, this 
assumption leads to worse description of data especially around 14 
MeV. In conclusion one can say that neither the uncertainty in the
level scheme at higher energies nor the level density parameters have
a significant effect on this analysis, based on the precise 
experimental data at low energies. Further analysis is required for 
the other reactions involved \cite{sudar96} in the study of this 
isomeric cross-section ratio especially at higher incident energies 
where the model calculation need additional improvement 
\cite{fessler98}.

\section*{ACKNOWLEDGEMENTS}
This work was supported in part by the Hungarian Research Found 
(Contract No. T 025024), the International Atomic Energy Agency, Vienna 
(Contract Nos. 7687/R0 and 8205), the International Science and
Technology Center (Project No. 176), and the Romanian Ministry of 
Research and Technology Grant No. 3028GR/B10.

\newpage

  TABLE I. Measured cross sections for the $^{58}$Ni$(n,p)^{58}$Co and 
$^{58}$Ni$(n,p)^{58}$Co$^m$ reactions, and measured and deduced 
isomeric cross section ratios for the former reaction.\\

{\tighten
\begin{tabular}{dcccc} \hline \hline
Neutron&   Measured   &   Measured   & Measured & Deduced\\ 
							   \cline{4-5}
energy & $\sigma(n,p)$&  $\sigma_m$  & 
                 \multicolumn{2}{c}{$\sigma_m/(\sigma_g+\sigma_m)$} \\
(MeV)  &     (mb)     &    (mb)      &                     \\ \hline
2.14   &	      & 19.2$\pm$1.3 & 0.260$\pm$0.014  \\
2.21   &	      &              & 0.259$\pm$0.011  \\
2.23   &	      &              & 0.274$\pm$0.026  \\
2.30   &	      & 20.6$\pm$1.4 & 0.234$\pm$0.010  \\
2.43   &	      &              & 0.254$\pm$0.012  \\
2.59   &	      & 	     & 0.277$\pm$0.011  \\
2.60   &	      & 29.1$\pm$2.0 &                  \\
2.74   &	      & 41.0$\pm$3.0 & 0.289$\pm$0.012  \\
2.83   &	      &              & 0.269$\pm$0.008  \\
2.84   &	      &  	     & 0.278$\pm$0.009  \\
2.94   &	      & 56.3$\pm$3.5 & 0.276$\pm$0.011  \\
10.3   &	      &	      	     & 0.445$\pm$0.019  \\
12.3   &	      &	      	     & 0.452$\pm$0.020  \\
13.4   &	      &	      	     & 0.545$\pm$0.004  \\
13.56  &413.6$\pm$13.2& 234$\pm$9    &             & 0.566$\pm$0.028\\
13.6   &	      &		     & 0.536$\pm$0.004  \\
13.74  &383.4$\pm$16.1& 218$\pm$9    &             & 0.569$\pm$0.034\\
13.96  &359.1$\pm$17.2& 197$\pm$8    &		   & 0.549$\pm$0.035\\
14.03  &	      &		     & 0.545$\pm$0.007  \\
14.05  &	      &		     & 0.552$\pm$0.005  \\
14.19  &329.9$\pm$12.8& 182$\pm$8    &		   & 0.552$\pm$0.032\\
14.42  &313.8$\pm$10.8& 170$\pm$7    &		   & 0.542$\pm$0.029\\
14.48  &	      &		     & 0.573$\pm$0.012  \\
14.61  &292.3$\pm$14.0& 173$\pm$11   &		   & 0.592$\pm$0.047\\
14.68  &	      &		     & 0.556$\pm$0.006  \\
14.78  &275.9$\pm$11.8& 150$\pm$6    &		   & 0.544$\pm$0.032\\
14.88  &	      &		     & 0.573$\pm$0.005  \\
\hline \hline
\end{tabular}
}

\newpage
TABLE II. The number of discrete levels $N_d$ up to excitation energy
$E_d$ used in Hauser--Feshbach calculations, taken from the 
corresponding references, and the low--lying levels as well as s--wave 
nucleon--resonance spacings $\overline{D}_{exp}$ in the nucleon energy 
range $\Delta$E above the corresponding binding energy B$_n$ 
\cite{wapstra85} used to obtain the BSFG parameters, i.e. the 
level--density parameter {\it a}, the ratio of the nuclear moment of 
inertia $I/I_r$, and the ground--state shift $\Delta$.\\

\begin{tabular}{cdccdcdcdcd} \hline \hline
Nucleus&$N_d$&$E_d$&Ref.&\multicolumn{4}{c}{Fitted level and resonance 
                                   data} & $a$ & $I/I_r$ & $\Delta$ \\
\cline{5-8}
 & &(MeV)& &$N_d$&$E_d$&$B_n+\frac{\Delta E}{2}$&$\overline{D}_{exp}$&
                                            (MeV$^{-1}$)&  & (MeV)  \\ 
 & &     & &    & (MeV) &        (MeV)          &   (keV)    \\ \hline
$^{59}$Ni& 13 & 1.948 & \cite{baglin93}&20 & 2.48 & 9.33 & 
					12.5$\pm0.9^a$&6.25 & 1.0 & 
							      -1.20 \\
\\
$^{58}$Ni& 28 & 4.475 & 
%\cite{peker90}
\cite{bhat97}
& 32 & 4.58 & & & 6.00 &1.0 & 
							       0.28 \\
\\
$^{58}$Co& 28 & 1.555 & 
%\cite{peker90}
\cite{bhat97}
& 28 & 1.56 & & & 6.60 &1.0 & 
							      -2.37 \\
$^{58}$Co& 28 & 1.555 & 
%\cite{peker90}
\cite{bhat97}
& 28 & 1.56 & & & 6.11 &0.5 & 
							      -2.40 \\
\\
$^{55}$Fe& 16 & 2.600 & \cite{junde91}& 16 & 2.60 & 9.55 & 
				  18.0$\pm2.4^a$&5.60 & 1.0 & -1.30 \\
\hline \hline
\end{tabular}
\bigskip

$^a$Reference \cite{vonach88}.

\newpage
\section*{Figure Captions}
FIG. 1. Comparison of the measured and calculated excitation
functions and isomeric cross section ratios for the 
$^{58}$Ni$(n,p)^{58}$Co$^{m,g}$ and $^{59}$Co$(n,2n)^{58}$Co$^{m,g}$
reactions. The calculated cross-section curves (c),(d) were obtained 
by using the fitted value of the branching ratio for the 
52.8 keV$\rightarrow$24.9 keV transition (solid curves), the isomeric 
cross-section ratios (a),(b) were found by using also the evaluated 
branching-ratio \cite{peker90} (dotted curves), using only the two
levels (dashed-dotted curves), as well as the fitted branching-ratio 
but considering two assigned 4$^+$ levels populating either the g.s. 
(lower dashed curves) or the isomeric state (higher dashed curves). 
For the experimental data see Refs. \cite{buczko95,sudar96}.

FIG. 2. (a) Partial cross sections for the compound nucleus 
formation versus the corresponding total angular momentum $J_{CN}$,
at the given incident energies of neutrons on the target nucleus
$^{58}$Ni. (b) The yrast plot for the residual nucleus $^{58}$Co,
of the adopted discrete levels \cite{peker90} including the spin 
assignment ($+$), while in the opposite case the spin values 
considered in the present calculations are additionally marked if the 
corresponding level has an adopted decay scheme ($\times$) or only the 
excitation energy ($\circ$); the yrast lines showed only for 
orientation correspond \cite{sh95,avr88} to the effective excitation 
energies obtained by using the BSFG parameters in Table II, and 
the nuclear moment of inertia for a rigid body  $I_r$ (dashed curve) 
and respectively one-half of $I_r$ (dotted curve) with a reduced 
radius $r_0$=1.25 fm.

FIG. 3. The same as Fig. 1(a), except the calculated values
are obtained by using (a) 28 discrete levels up to the excitation
energy $E^*$=4.475 MeV of the nucleus $^{58}$Ni, while for the nucleus
$^{58}$Co are used either 28 discrete levels \cite{bhat97} (solid 
curve) or 29 levels \cite{peker90} (dashed curve) up to $E^*$=1.555 
MeV, 3 discrete levels up to $E^*$=0.053 MeV (dotted curve), or 2 
discrete levels up to $E^*$=0.025 MeV (dashed-dotted curve), and (b) 
only 2 discrete levels up to $E^*$=0.025 MeV for the nucleus $^{58}$Co 
while for the nucleus $^{58}$Ni are used either 28 levels up to 
$E^*$=4.475 MeV (solid curve), only g.s. (dashed curve), 2 levels up 
to $E^*$=1.454 MeV (dashed-dotted curve), or 4 levels up to 
$E^*$=2.776 MeV (dotted curve).
\end{document}